# ВНЕГАЛАКТИЧЕСКИЕ ТЭВ-НЫЕ ФОТОНЫ И ПРЕДЕЛ СПЕКТРА НУЛЕВЫХ КОЛЕБАНИЙ

Жогин И. Л.[a],[1]

[a] ИХТТМ, г. Новосибирск, 630090, Россия.

Вселенная не совсем прозрачна для фотонов очень высокой энергии, превышающей 100 ГэВ, из-за их поглощения межгалактическим фоновым ИК-излучением, с образованием электрон-позитронных пар. Ряд наблюдений таких фотонов от источников, находящихся за пределами нашей Галактики, указывают на возможную аномалию – неожиданно низкое поглощение ТэВ-ных фотонов. Высказывались предположения, что аномалия может быть связана с эффектами "новой физики", например, аксионоподобными частицами, либо с возможным нарушением Лоренц-инвариантности.

Здесь предлагается иное объяснение, что аномалия есть проявление границы спектра нулевых колебаний. Предположено, что эта граница $U_{ZV}$ изотропна в системе отсчета, где изотропно реликтовое излучение, и получена оценка: $U_{ZV} \approx 7.4$ ТэВ. Отмечено, что наличие $ZV$-границы приводит также к увеличенному времени бета-распада ускоренных частиц с Лоренц-фактором $\gamma > 50$ (в дополнение к обычному $\gamma\tau^{(\beta)}$).

Распространено мнение, что спектр нулевых колебаний продолжается вплоть до планковской энергии (в естественных единицах, гравитационная постоянная связана с квадратом планковской длины). Существует, однако, 5D-вариант теории Абсолютного Параллелизма (АП), свободный от сингулярностей решений, где появляется большая характерная длина $L$, определяющая толщину расширяющейся сферической $S^3$-оболочки (космологическое решение как продольная волна, бегущая по радиусу) в сопутствующей системе. Ньютоновский закон $\sim 1/r^2$ сменяется на $1/r$ на расстояниях, превышающих $L$, а планковская длина (составная величина) "возникает" из $L$ при выборе традиционного масштаба энергии-импульса (в котором энергия фотона есть его угловая частота).

Отмечаются особенности данной теории – классификация 15-ти поляризаций (поляризационных мод), тензор энергии-импульса (в продолженных уравнениях 4-го порядка), топологические заряды и квазизаряды локализованных конфигураций поля.

*Ключевые слова*: нулевые колебания; внегалактические ТэВ-ные фотоны; планковская длина; Абсолютный Параллелизм.

# EXTRAGALACTIC TEV PHOTONS AND THE ZERO-POINT VIBRATION SPECTRUM LIMIT

Zhogin I. L.[a],[1]

[a] ISSCM, Novosibirsk, 630090, Russia.

There are observations indicating a possible anomalous transparency of intergalactic space (filled with infrared background light) for extragalactic gamma-rays of very high energy ($> 100$ GeV). The anomaly is usually associated with effects of some new physics.

However, another explanation is possible — as a manifestation relating to a cut-off of the zero-point vibration spectrum. It is assumed that this boundary $U_{ZV}$ is isotropic in the reference frame, where the cosmic microwave background (CMB) radiation is isotropic, and an estimate is obtained: $U_{ZV} \approx 7.4$ TeV. It is noted that the presence of a $ZV$ boundary also leads to an increased beta decay time of accelerated particles with the Lorentz factor $\gamma > 50$ (in the CMB rest frame; in addition to the usual $\gamma\tau^{(\beta)}$).

It is widely believed that the ZV-spectrum continues up to the Planck energy (in natural units, the gravitational constant is related to the square of the Planck length). There is, however, a 5D variant of the Absolute Parallelism theory (AP), free from singularities of solutions, where a large characteristic length $L$ appears, which determines the thickness of expanding spherical $S^3$ shell (a cosmological solution as the longitudinal wave along the radius) in co-moving co-ordinates. Newton's Law $\sim 1/r^2$ is replaced by $1/r$ at distances exceeding $L$, and the Planck length (a composite parameter) "arises" from $L$ when switching to the conventional energy-momentum scale (where the energy of a photon is its angular frequency).

The theory features are briefly exposed – description of 15 polarizations (degrees of freedom), the energy-momentum tensor (in prolonged 4th order equations), topological charges and quasi-charges of localized field configurations.

### Introduction

Very-high-energy (VHE) gamma-rays, $E_\gamma > 100\,\text{GeV}$, are recorded by ground-based observatory facilities, clusters of atmospheric Cherenkov telescopes, etc. Extragalactic sources of VHE photons are active galactic nuclei, such as blazars (Markarian 501 [1], quasar 3C 279 [2]), while within the Galaxy VHE photons are produced, for instance, by the Crab Nebular pulsar (2 kpc; $E_\gamma$ over 100 TeV). The LHAASO observatory reported photons with energies of $1\ldots1.4$ PeV; it is possible that some of these quanta came from outside the Galaxy [3].

The universe is not entirely transparent to such hard photons, as they are absorbed by the extragalactic background light (EBL, which includes also photons with energies $E_b = 0.01\ldots4\,\text{eV}$ in addition to the cosmic microwave background radiation, CMB) through the electron-positron pair production. The threshold depends on the electron mass $m_e$, $E_\gamma\,E_b > m_e^2$, and the cross-section (absorption) peaks [3] if

$$E_\gamma\,E_b \approx 1\ldots5\times10^{12}\,\text{eV}^2. \tag{0.1}$$

Both brightness of VEH-sources and their limiting energies $E_\gamma$ can increase dramatically during **_flares_**.

Let us consider a few sources of VHE photons, with their redshift $z$, distance $L$, and energy limit $E_\gamma$. The distance is estimated via the expression (we assume the linear expansion model $a(t)\propto t$)

$$L = c\,t_0\,z/(1+z),\ \ L[\text{Mpc}] = 4283\,z/(1+z)\,;$$

that is, we use $H_0 = t_0{}^{-1} = 70\ \text{km}\,\text{s}^{-1}\text{Mpc}^{-1}$.

Here are just three such sources:

⊛$^a$ the Mkn 501 blazar [1] (HEGRA), $z = 0.0336$, $L = 140\,\text{Mpc}$, $E_\gamma = 20\,\text{TeV}$;

⊛$^b$ the 3C 279 radio-quasar [2] (MAGIC), $z = 0.536$, $L = 1.495\,\text{Gpc}$, $E_\gamma = 0.3\ldots0.5\,\text{TeV}$;

⊛$^c$ GRB 221009A [4,5] (LHAASO / Carpet-2), $z = 0.1505$, $L = 560\,\text{Mpc}$, $E_\gamma = 18\,\text{TeV}$ / $251\,\text{TeV}$.

## 1. TeV gamma-ray crisis?

The gamma-ray burst (GRB) of October 9th, 2022, had a record-breaking brightness [4,5]; details on the Carpet-2 facility recording a 251 TeV photon were reported at the workshops of Theoretical Physics Department of the Institute for Nuclear Research [4] (S. Troitskiy, V. Romanenko; there are some problems with this photon: the proximity of Galactic disk and the presence of 2–3 marginal muons).

The plots on Figure 1 illustrate the mean free path of VHE photons along with the spectra of EBL and the Mkn 501 blazar (taken from [1]).

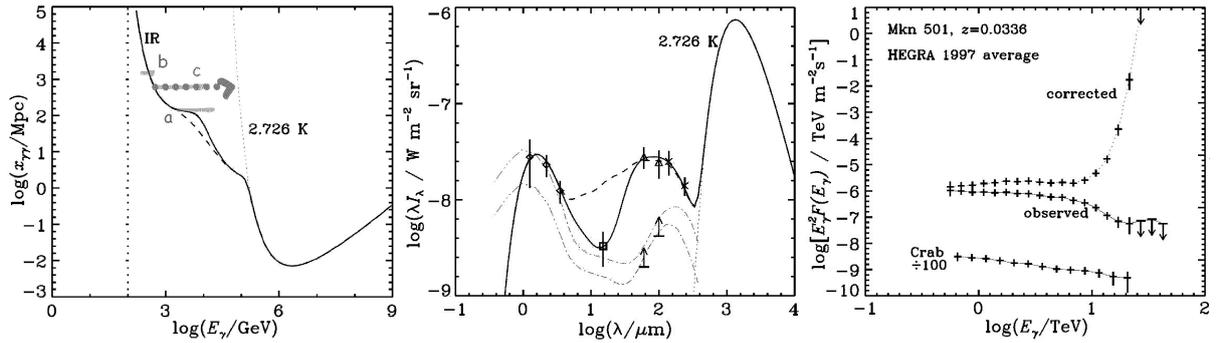

**Fig. 1.** The mean free path of VHE photons, the EBL intensity, and the Mkn 501 spectrum correction [1].

New measurements are being made, the spectra of background light (EBL) and TeV sources are being discussed; however, many authors believe that the extragalactic background light (EBL) is anomalously transparent for TeV photons, cf. Figure (the corrected Mkn 501 spectrum), and that the anomaly explanation requires a certain new physics [1–6] (such as axion-like particles [3] or models violating the Lorentz invariance [6]).

It is simpler, however, to connect this anomaly to a manifestation of the zero-point vibration spectrum limit, $U_{ZV}$. (It is hardly possible to stretch such a good thing as zero vibrations to infinity.)

One can assume that the ZV-ensemble is isotropic in the coordinate system, where the CMB radiation is almost isotropic (say, with an accuracy of about $\sim 10^{-5}$, or $v \sim \pm 3$ km/s).

## 2. The limit of zero-point vibration (ZV) spectrum

An unstable particle (with a lifetime $\tau_0$), whose decay is associated with ZVs of the energy scale $U_0$, when moving relative to the ZV + CMB "ether" with the Lorentz factor $\gamma_e$ will sense this ZV spectrum boundary (i.e., the ZVs weakening in the backward direction) and will live some longer than mere $\gamma_e \tau_0$, if

$$U_0 \geqslant U_{ZV}/(2\gamma_e), \ \text{ or } \ U_{ZV} \leqslant 2\gamma_e U_0. \tag{2.1}$$

The photons with $E_\gamma = 16$ TeV and $E_b = 0.3$ eV form $e^+e^-$ pairs in a zero-momentum frame with the Lorentz factor (see Eq. (0.1); this simple estimate is for the case of a head-on collision)

$$\gamma_e^{(p)} \approx 0.5\sqrt{E_\gamma/E_b} \approx 3.7{\cdot}10^6,$$

and the ZV energy required for pair production is about $U_0^{(p)} \approx 10^6$ eV (it is the $e^+e^-$ pair mass). If we assume that the ZV anomaly is already coming in effect, then the next estimate follows from Eq. (2.1):

$$U_{ZV} \approx 2\gamma_e^{(p)} U_0^{(p)} \approx 7.4 \text{ TeV}. \tag{2.2}$$

The shortest (or the "heaviest") ZVs are seemingly involved in the weak interactions. Therefore it would be very interesting to measure the anomalous increase in lifetime (compared to $\gamma\tau_0$) for particles featuring $\beta$-decay. Given that $U_0^{(\beta)} \approx 80$ GeV (the $W^\pm$ boson mass) and using Eq. (2.2), it is possible to estimate the Lorentz factor (relative to the ZV "ether") of the anomaly onset:

$$\gamma_e^{(\beta)} = U_{ZV}/(2U_0^{(\beta)}) \approx 46. \tag{2.3}$$

Most likely, as is typical for the weak interactions, this lifetime anomaly ($\tau^{(\beta)}$-anomaly) should differ for particles of different helicity.

In addition to muons [the idea of $\mu^\pm$-collider (Budker, Skrinsky, *etc.*) is advancing somewhat, see MICE.iit.edu], $\beta^\pm$-decaying nuclides such as $^3$H ($\tau_{\beta-} = 12.3$ y) and $^7$Be ($\tau_{\beta+} = 53$ d) are of special interest. One should note that $(u, d)$-quarks already have Lorentz factors $\gamma_q$ about $35\ldots70$ in their nucleons, and it is very significant – cf. Eq. (2.3).

(For the bottle-beam neutron anomaly [7], the velocity of thermal neutrons $v_{beam}$ is too low; but bottle-neutrons often come in contact with the wall nucleons (protons), while their $d$-quark velocities ('relativism') can decrease — as can the lifetime.)

It is generally accepted that the ZV spectrum should be extended till the Planck energy. There exists, however, a 5D theory [8–10] in which the Planck length $\lambda_{Pl}$ is a composite parameter that does not correspond to any characteristic scale, and where gravity does not have to be quantized.

## 3. Periodic (annual and diurnal) changes in beta decay rates

Several experiments yielded evidence for the variability of beta decay rates (a number of nuclides were involed) [11]; the amplitude of annual oscillations is of the order $10^{-3}$, or 0.1%. The situation is still rather controversial because environmental influences could be in effect (along with other issues [12]).

Some experiments reported about diurnal variations in beta decays [13].

The Earth orbital velocity is about 30 km/s, and it adds to or subtracts from the Sun velocity relative to the CMB rest frame (sure one should account the ecliptic slope), $v_{\odot} \approx 369.8$ km/s; it corresponds to annual disturbances of the quark Lorentz factor $\gamma_q \left(1.0012 \pm 10^{-4}\right)$ – quite a small variation.

The direction of $\vec{v}_{\odot}$ (to Leo/Crater) has the next coordinates in the second equatorial system [14]:
- ⊙  the right ascension $\alpha = 167°.942 \pm 0°.007$
- ⊙  and declination $\delta = -6°.944 \pm 0°.007$ (J2000);
- ⊙  the galactic coordinates are $(l, b)[\deg] = (264.02, 48.25)$.

An experimental setup can carry a peculiar vector, $\vec{v}_p$, e.g, directed from the source to detector, and the rate of beta decays could slightly depend on the angle between these two vectors. So, variations of this angle due to the Earth rotation can cause diurnal variations of beta decays in that experiment.

The 'directionality hypothesis' is also considered [12, 13]; usually the Sun direction is regarded as special.

## Conclusion

Perhaps in order to achieve 0.1 scale $\tau$-anomaly, there would be enough to accelerate tritons, the lightest beta decaying nuclides, to moderate speeds $v/c \sim 0.1$ (i.e., the triton momentum is about 0.3 GeV). This kind of experiment, where particles will collide not with other particles, but with the 'ZV-ether', will allow us to ask Nature new questions.

Special relativity (SR) united space and time but did not explain existence of any field or particle. General relativity (GR) relates gravity to space-time curvature; the other fields/particles form the energy-momentum tensor, EMT, and remain unexplained.

Einstein wasn't content with GR (the complete and true theory should explain more); he compared the GR-equation sides with a marble palace (the LHS, Einstein's tensor $G_{\mu\nu} = R_{\mu\nu} - g_{\mu\nu} R/2$, $G_{\mu\nu;\nu} \equiv 0$) and an old shed (the RHS with EMT, $T_{\mu\nu}$).

Later Einstein explored the co-frame field $h^a{}_\mu(x^\nu)$, with the metric $g_{\mu\nu} = \eta_{ab} h^a{}_\mu h^b{}_\nu$ where $\eta_{ab}$ is Minkowski's metric, and second order equations which symmetry unites symmetries of both SR (Latin indexes) and GR (Greek ones) – the third (or united) relativity, known as Absolute Parallelism (AP).

The list of compatible $2^d$-order AP equations (found by A. Einstein and W. Mayer in [15]; they used D=4) includes the two-parameter class of Lagrangian equations and three more classes. And there exists the exceptional equation (EE), non-Lagrangian, which solutions don't allow co-singularities (the principal terms do not remain regular for degenerate co-frame matrices), and, if D=5, contra-singularities (related to degenerate contra-frame densities of some weight) [9].

The additional spatial dimension manifests itself both in the cosmological expansion (there are spherically symmetric non-stationary solutions as a longitudinal wave running along the radius and forming a cosmological shallow waveguide, a region with non-zero Ricci tensor), and also in the nonlocal behavior of elementary particles (large size along the extra dimension; localized configurations of the frame field can carry discrete information – topological charges and, if configurations have some symmetry, topological quasi-charges).

It seems the frame field is only twice as large in number of components: ($D^2$–D) compared to vacuum GR. However, the increase in the number of polarizations (polarization modes or degrees of freedom, PDF) is more pronounced: D(D–2)=15 compared to D(D–3)/2 =5, the number of GW-polarizations in D=5 (two usual tensor plus additional three vector GW polarizations).

A simple (maybe the simplest) compatible AP equation (2d order; non-Lagrangian) looks as follow:

$$\Lambda^a_{\ \mu\nu;\lambda} g^{\nu\lambda} = 0, \ \text{where} \ \Lambda^a_{\ \mu\nu} = h^a_{\ \mu,\nu} - h^b_{\ \nu,\mu} = 2h^a_{\ [\mu;\nu]};$$

together with the identity $\Lambda^a_{\ \mu\nu;\lambda} \equiv 0$ ($\Lambda$-identity) it looks after linearization as a D-fold Maxwell equation, so the number of polarizations is D(D–2)=15; D=5 is the must for the EE which has the same number of polarizations as the simple equation.

These 15 polarizations can be separated [10] on four classes according to their very different amplitudes (and functions; a higher class means many orders smaller amplitudes) as they relate to various irreducible parts of tensor $\Lambda$ (and it's derivative $\Lambda'$) such as:

⊙  $\Phi_\mu = h^a_\mu \Lambda^a_{\ \nu\mu}$ (3+1 polarizations, $2^d$- and $1^{st}$-class; no gradient symmetry);

⊙  $S_{\mu\nu\lambda} = 3\Lambda_{[\mu\nu\lambda]}$ (3 polarizations, $1^{st}$-class);

⊙  and the Riemannian curvature tensor (or the Weyl tensor; 5 polarizations, $3^d$-class);

⊙  three unstable pol-ns ($0^{th}$-class) grow linearly under action of three stable polarizations relating to $f_{\mu\nu} = 2\Phi_{[\mu;\nu]}$ ($2^d$-class), while $h'^2$-terms are tiny (the divergence of $\Lambda$-identity):

$$\Lambda_{\lambda\mu\nu;\tau;\tau} = -\frac{2}{3} f_{\mu\nu;\lambda} + (\Lambda\Lambda', \Lambda^3) \ (\Box S \approx \Box\Phi \approx 0 \approx \Box f \approx \Box Riem - \text{stable polarizations}).$$

Only very small $2^d$-class polarizations take part in the energy-momentum tensor which apear in the prolonged 4th order equation (symmetrical part; 4th order gravity); it follows also from a Lagrangian quadratic in 2d order field equations (or the *weak* Lagrangian – in Ibragimov's sense).

Non-stationary $O_4$-symmetrical solutions exist which resemble a (single) longitudinal wave in Chaplygin gas [9] (the $1^{st}$-class pol-n relating to $\Phi_\mu$; others don't survive in this symmetry); the wave can serve as a cosmological shallow waveguide for tangential shorter waves, with ultrarelativistic expansion and different evolution of waves' amplitudes – according to the structures of quadratic terms (whether they include $0^{th}$-class parts or only lower class ones).

Non-linear localised $h$-field configurations can carry digital information – topological charges and (for symmetrical configurations) quasi-charges (when $0^{th}$-class waves become large enough), and a QM-like 4D-phenomenology emerges through averaging along the huge extra-dimension, along a length $L$, the width of large-scale $O_4$-wave in co-moving coordinates [8,9]; note, two thin lines in a 4d-space have tiny chances to intersect in a single approach. The complete description is five-dimensional, not four!

Finally, it is useful to introduce auxiliary 4D-fields (quantised avatar-fields) for phenomenological description of topological (quasi)particles prone to interact (a kind of new actors!). So the overall picture turns out to be complex and interesting, and many features of the Standard Model becomes understandable including the lepton flavours and neutral (perhaps $CP$-symmetrical, like photons) neutrinos.

**Авторы**

**Жогин Иван Львович**, к.ф.-м.н., ИХТТМ СО РАН, ул. Кутателадзе, д. 18, г. Новосибирск, 630090, Россия.
E-mail: zhogin@mail.ru

**Authors**

**Zhogin Ivan L'vovich**, Ph.D., ISSCM SB RAS, Kutateladze st., 18, Novosibirsk, 630090, Russia.
E-mail: zhogin@mail.ru